# Thin Film Magnesium Boride Superconductor with Very High Critical Current Density and Enhanced Irreversibility Field

C.B. Eom[†§], M.K. Lee[†], J.H. Choi[†§], L. Belenky[†], X. Song[§], L.D. Cooley[§], M.T. Naus[§], S. Patnaik[§], J. Jiang[§], M. Rikel[§], A. Polyanskii[§], A. Gurevich[§], X.Y. Cai[§], S.D. Bu[†], S.E. Babcock[†§], E.E. Hellstrom[†§], D.C. Larbalestier[†§] N. Rogado*, K.A. Regan*, M.A. Hayward*, T. He*, J.S. Slusky*, K. Inumaru*, M.K. Haas* and R.J. Cava*

[†] Department of Materials Science and Engineering, University of Wisconsin, 1509 University Avenue, Madison, WI 53706 USA
[§] Applied Superconductivity Center, University of Wisconsin, 1500 Engineering Drive, Madison, WI 53706 USA
[*] Department of Chemistry and Princeton Materials Institute, Princeton University, Princeton, NJ 08544 USA

**The discovery of superconductivity at 39 K in magnesium diboride [1] offers the possibility of a new class of low-cost, high-performance superconducting materials for magnets and electronic applications. With twice the critical temperature of $Nb_3Sn$ and four times that of Nb-Ti alloy, $MgB_2$ has the potential to reach much higher fields and current densities than either of these technological superconductors. A vital prerequisite, strongly linked current flow, has already been demonstrated even at this early stage [2-5]. One possible drawback is the observation that the field at which superconductivity is destroyed is modest. Further, the field which limits the range of practical applications, the irreversibility field $H^*(T)$, is ~7 T at liquid helium temperature (4.2 K), significantly lower than ~10 T for Nb-Ti [6] and ~20 T for $Nb_3Sn$ [7]. Here we show that $MgB_2$ thin films can exhibit a much steeper temperature dependence of $H^*(T)$ than is observed in bulk materials, yielding $H^*$(4.2 K) above 14 T. In addition, very high critical current densities at 4.2K, 1 $MA·cm^{-2}$ at 1 T and $10^5$ $A/cm^2$ at 10 T, are possible. These data demonstrate that $MgB_2$ has credible potential for high-field superconducting applications.**

Three $MgB_2$ films were prepared by pulsed laser deposition at room temperature under 0.3 Pa of argon from a sintered $MgB_2$ target using a Lambda Physik KrF laser (248 nm) operating at 4 $J/cm^2$ and 10 Hz onto (111) oriented single crystal $SrTiO_3$ substrates. Before each deposition, the vacuum chamber was pumped to a base pressure of $2 \times 10^{-7}$ Torr in order to minimize the oxidation during the deposition. The $MgB_2$ target was prepared as described previously [2, 8]. After deposition, Film 1 was annealed in an evacuated niobium tube (0.6 cm diameter × 5 cm long) at 850 °C for 15 minutes. Film 2 was annealed in an evacuated quartz tube (1.7 cm diameter × 15 cm long) at 750 °C for 30 minutes and quenched to room temperature. Film 3 was annealed in a tantalum envelope inside an evacuated niobium tube at 750 °C for 30 minutes. Magnesium pellets were included in each tube to prevent magnesium loss. The film thickness was approximately 5000 Å.

The films' chemical composition was determined by wavelength dispersive spectroscopy (WDS) in an electron microprobe at 3 keV acceleration voltage to avoid x-ray intensity from the substrates. Before annealing, the films had a Mg:B:O atomic ratio of about 1.0:1.0:0.17, the oxygen probably coming from the target or background gas during deposition. The primary differences in the films are the measured ratios of Mg:B:O obtained after annealing. Film 2 had a ratio of ~1.0: 0.9: 0.7, which corresponds to a relative oxygen concentration approximately twice that of film 1 and 3 and higher Mg content. By contrast, film 1 and film 3 annealed in small-volume Nb and Ta tubes show 1.0:1.0: 0.4 and 1.0: 1.2: 0.3 ratios, respectively. Film 1 has a dark spot in the center of the sample, which is excluded from our chemical analysis, but could not be excluded from the 3 × 3 mm squares used for superconducting property characterization. Samples 2 and 3 exhibited no similar inhomogeneity. We believe that the higher oxygen of post-annealed film 2 comes from the relatively poor vacuum (approximately 150 Pa) and the large volume of the quartz tube. As shown later, film 2 has dramatically different superconducting properties compared to films 1 and 3.

The critical temperature $T_c$ was determined by both inductive SQUID magnetometer and resistive measurements, which showed principal magnetic transitions at 28 to 32 K. Film 1 had a small amount of superconducting material up to 38.5 K. The resistive transitions, shown in Fig. 1, have onset values of 36, 35, and 30 K for films 1, 2, and 3 respectively, and fall to zero resistance just above the principal inductive transition. The room temperature resistivity $\rho$(300 K) of film 2, 390 $\mu\Omega·cm$, is clearly much higher than that for film 1 (125 $\mu\Omega·cm$) and film 3 (60 $\mu\Omega·cm$). The uncertainty of the resistivity value due to the large voltage contacts relative to sample dimension is about ±20%. High-resistivity film 2 also exhibits a slight resistivity increase with decreasing temperature, indicating a resistivity ratio, RR = $\rho$(300 K) / $\rho$(40 K), less than 1. By contrast, the RR of film 1 is about 1.5, while that of film 3 is 1.3. For comparison, the RR of Nb-Ti alloys (Ginzburg-Landau parameter $\kappa \approx 40$) is seldom higher than 1.4 [6], while that of $Nb_3Sn$ ($\kappa \approx 30$) is typically between 2 and 5 [7]. Good bulk samples of $MgB_2$ can attain RR values of up to 25 and have $\kappa \sim 25$ [4, 5].

Magnetization curves were obtained using 3 mm × 3 mm squares at 4.2-30 K and fields up to 14 T in a vibrating sample magnetometer (VSM). The 4.2 K data shown in fig. 2a demonstrate important differences between film 2 and films 1 and 3 and also between the films and bulk samples. Hysteresis curves close at fields very much higher for film 2, ~14 T, than for film 1 (~6 T) and film 3 (~6.5 T). We were not able to unambiguously determine $H_{c2}$ due to the smooth blending of the superconducting magnetization into the background magnetization of substrate and sample holder. This is in marked contrast to the behavior seen for bulk samples, which have a significant reversible contribution to the remaining hysteretic magnetic moment above the primary loop closure field [2,4,9-11]. The O-rich film 2 has a much higher irreversibility field than films 1 and 3 up to 25 K.

Magneto-optical examination of the films showed strong flux shielding with only slightly distorted roof patterns, indicative of macroscopic supercurrents flowing over the whole films. This validates use of the standard expression [12] for the magnetization of a thin film, $J_c = 30·\Delta m·V^{-1}·r^{-1}$, to extract critical current densities $J_c(H,T)$, where $\Delta m$ is the width of the hysteresis loop, $V \approx 4.5 \times 10^{-6}$ cm$^3$ is the film volume, and $r = 0.15$ cm is the sample half-width. Figure 2b shows that the highest $J_c$ value at 1 T, 4.2 K exceeds 3 MA·cm$^{-2}$ for the low-resistivity film 3. However, the much larger $H^*(4.2$ K$)$ of film 2 enables $J_c$ to achieve 1 MA·cm$^{-2}$ at 1 T and to remain above $10^5$ A·cm$^{-2}$ up to 10 T, well above $H^*$ for films 2 and 3. Film 2 also exhibits substantial high-field current densities up to 20 K. $J_c$ values of film 1 are well below those of film 3 at 4.2 and 20 K, which may be due to its inhomogeneity.

Bulk pinning-force curves, $F_p(H) = \mu_0 H \cdot J_c(H)$ are shown in fig. 2c. Very strong flux pinning, in excess of 40 GN·m$^{-3}$ at 4.2 K, is reached by film 3, while $F_p$ approaches 20 GN·m$^{-3}$ at 5 T for film 2 and remains above 10 GN·m$^{-3}$ between 1 and 10 T. These data are comparable to or exceed the pinning force of the established technological superconductors Nb47wt.%Ti [6] and Nb$_3$Sn [7], which lie in the range 15-30 GN·m$^{-3}$.

Fig. 3 summarizes the irreversibility fields for bulk and thin film MgB$_2$. The gray bands indicate established $H_{c2}(T)$ and $H^*(T)$ lines for bulk MgB$_2$, based on inductive [2,4,5,9-11] and resistive [4,5,9,11] measurements. The extrapolation to zero of the Kramer function ($J_c^{0.5} H^{0.25}$) was used to define the irreversibility field. Plots for films 1 and 3 were nearly straight, just as in bulk samples [2]. Kramer plots for film 2 were curved because of stronger pinning at higher fields. The $H^*(T)$ line for film 2 also has a much steeper slope, $dH^*/dT$ 0.7 T·K$^{-1}$, than the $H^*(T)$ data for bulk samples, which have a slope of about 0.3 T·K$^{-1}$. This steeper $H^*(T)$ characteristic allows $H^*(T)$ for film 2 to intersect and exceed bulk sample $H_{c2}(T)$ values below about 10 K. In strong contrast to the film 2 data, the $H^*(T)$ data for lower resistivity films 1 and 3 lie along the bulk $H^*(T)$ line, except for a downturn at higher temperature that reflects their lower critical temperature (30-32 K).

Transmission electron microscopy (TEM) imaging and selected area diffraction pattern (SADP) analysis from widely spaced regions were used to characterize the phase constitution and microstructure of film 2. Each region had an area of ~1 μm$^2$ in the plane of the film. All of the SADPs were ring-pattern types (Fig. 4), a characteristic of fine-grained material, and indicated the presence of significant volume fractions of both MgB$_2$ and MgO. All MgB$_2$ rings from crystal planes other than ($hk0$) were absent when the electron beam was directed along the film surface normal. Arcs, rather than rings, appeared at the MgB$_2$ interplanar spacings when the beam was tilted significantly away from the film normal. These two characteristics of the SADPs show that the MgB$_2$ has a crystallographic texture, in which most grains have their $c$-axis oriented along the film surface normal but have no preferred orientation in the plane of the film. TEM imaging showed that the MgB$_2$ grain size was remarkably small, ~10 nm (Fig. 4). As in the WDS microprobe results, energy-dispersive TEM x-ray microanalysis suggests a considerably higher concentration of O compared to what we have observed in TEM specimens from bulk MgB$_2$ [13].

All three films were probed by x-ray diffractometry using both a point and a two-dimensional area detector to collect the diffracted intensity and probe its spatial distribution. All films showed a [00$l$] texture with a large mosaic spread normal to the substrate and random in-plane orientation, consistent with the SADP shown in Fig. 4. The rocking curve widths of the (002) reflection of MgB$_2$ are 4.8°, 7.3° and 8.2° for films 1, 2 and 3, respectively. Furthermore there is significant broadening of the (002) x-ray peaks in 2θ scans. Using the Sherrer formula for the determination of the grain size $d$ from the broadening of x-ray diffraction peaks, $d = 0.9 \lambda/B \cos\theta$ where $B$ is the full width at half maximum of the 2θ diffraction peaks, grain dimensions along the $c$-axis are 17 nm, 7 nm and 8 nm for films 1, 2 and 3, respectively. To within the limit of the analysis, the grain size of film 2 measured by TEM and XRD are the same. All are very small, though the grain size of film 1 is significantly larger. The MgO grains appeared to be randomly oriented with a grain size of ~10 nm in all three films.

X-ray diffraction studies also revealed that the MgB$_2$ $c$-lattice parameter is larger in film 2 than in either film 1 or film 3 and also in bulk samples. Fig. 5 presents θ-2θ scans for films 1, 2, and 3, which clearly show that the (002) MgB$_2$ peak of film 2 is shifted to lower angle. The c-lattice parameter of MgB$_2$ in film 2 is 0.3547±0.0007 nm, substantially larger than the 0.3521 nm for film 1 and 3, which is identical to the bulk value [1,3]. (The (111) and (222) peaks of SrTiO$_3$ were used as a reference in these calculations). We believe the expansion of the c-lattice parameter in MgB$_2$ is due to oxygen alloying in boron layer.

Taken as a whole, the experiment suggests that the thin film process can alloy magnesium diboride. Unlike bulk samples, which appear to have a common $T_c$, $H_{c2}(T)$, and $H^*(T)$, here we see a strong enhancement of $H^*(T)$ that correlates well to the strongly enhanced resistivity of film 2 and to concomitantly reduced electron mean free path and coherence length. Films 1 and 3, which were annealed in metal tubes and have lower oxygen, have metallic resistivity and $H^*(T)$ values close to the now well established bulk-sample line. Film 2, which was annealed in quartz and has higher measured oxygen and an enhanced c-axis parameter (0.3547 nm versus 0.3521 nm for bulk) has a weak increase of resistivity with decreasing temperature and a much steeper $H^*(T)$ curve, suggesting that oxygen can substitute for boron and expand the c-axis of MgB$_2$. In both alloy and intermetallic-compound superconductors, substitutional alloying to decrease the electron mean free path is very important for enhancing high-field superconductivity, as indicated by the direct dependence of $H_{c2}(0)$ on the normal-state resistivity through $H_{c2}(0) = 3.11 \rho \gamma T_c$ [14] ($\gamma$ is the coefficient of electronic specific heat). Taking the value of $\gamma = 153$ J·m$^{-3}$·K$^{-2}$ for MgB$_2$ [15], the resistivity values predict $\mu_0 H_{c2}(0)$ values of 12, 57, and 7.6 T for films 1, 2, and 3 respectively, assuming the dirty limit applies. Although we were not able to deduce $H_{c2}$ values from the magnetization measurements, these predictions are at least qualitatively reasonable given the data in fig. 3, which suggest extrapolated values of $\mu_0 H^*(0)$ of 10, 22, and 7 T for films 1, 2 and 3, respectively. Moreover the very high $J_c$ values indicate strong flux pinning and confirm the absence of electromagnetic granularity [2]. The source of flux pinning is not obvious in the TEM examination of the grains, but the exceptionally fine 10 nm-scale of the grain size would be consistent with strong grain boundary pinning, as in Nb$_3$Sn [16]. In summary, it appears that MgB$_2$ can be alloyed, probably with oxygen, so as to increase the irreversibility field and the flux-pinning properties to levels where they may compete with today's materials of choice, Nb47wt.%Ti and Nb$_3$Sn.

**Acknowledgments**

This work was supported by funding from the United States Department of Energy, the Air Force Office of Scientific Research, the National Science Foundation through the Materials Research Science and Education Center for Nanostructured Materials and David Lucile Packard Fellowship (CBE). The work at Princeton was supported by the National Science Foundation and the U.S. Department of Energy.

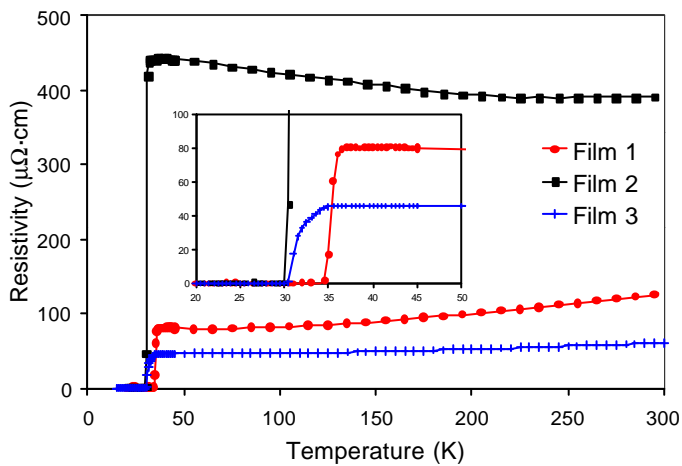

Figure 1: Resistivity as a function of temperature. Film 1 is indicated by the red curve with crosses, film 2 by the black curve with squares, and film 3 by the blue curve with circles. The inset shows a magnified view of the resistive transitions near the critical temperature.

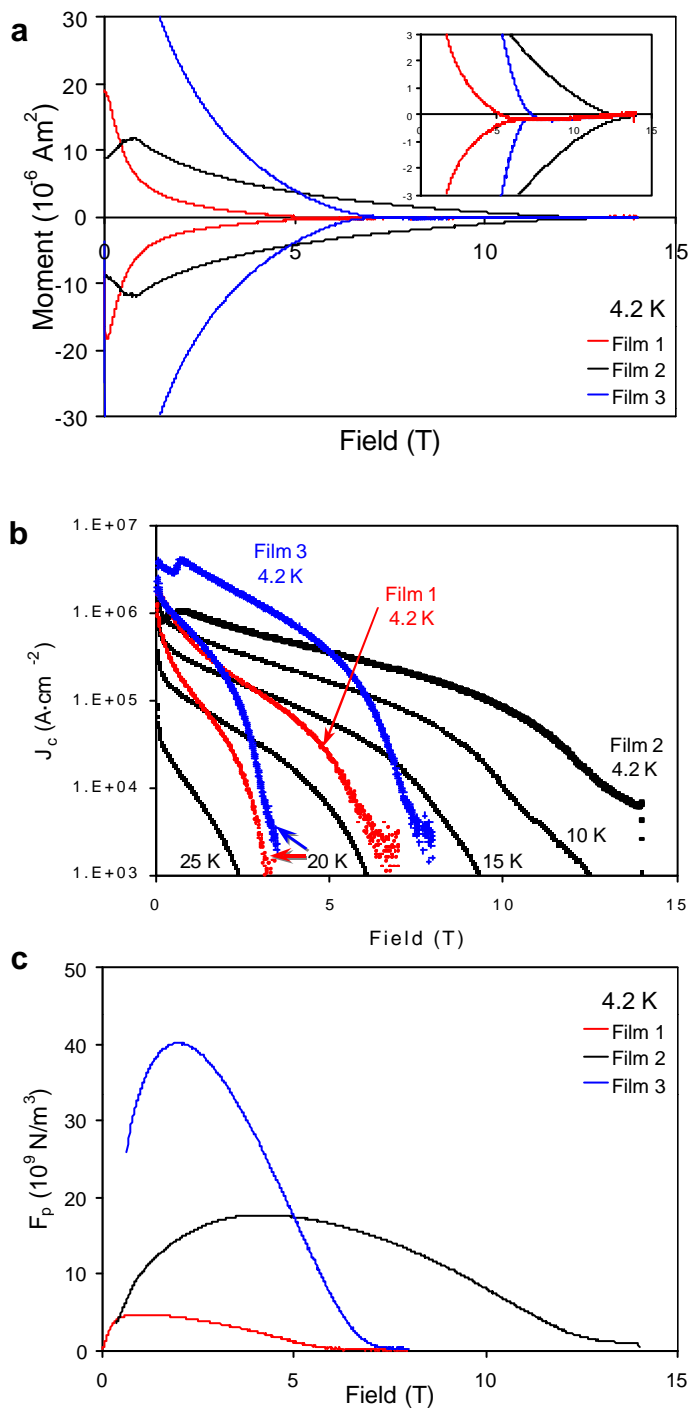

Figure 2: Magnetization measurements for films 1 (red data and curves), 2 (black), and 3 (blue). In (a), raw magnetization data, obtained using a vibrating sample magnetometer, is compared at 4.2 K. The inset shows a magnified view of the closure of hysteresis loops, after subtracting a linear background due to the paramagnetic moment of the samples and the sample holder. In (b), the derived values of the critical current density are plotted as a function of field at 4.2 and 20 K for films 1 and 3, and at 4.2, 10, 15, 20, and 25 K for film 2. Notice that the data for film

2 at 4.2 K are above $10^5$ A·cm$^{-2}$, a common benchmark for superconducting magnet applications, nearly to 10 T. In (c), the bulk pinning forces $F_p(H) = J_c(H)\cdot\mu_0 H$ at 4.2 K are shown. Inhomogeneous film 1 has lower $J_c$, while the more optimum films 2 and 3 show very strong flux pinning, comparable to the range of $J_c$ values found for technological superconductors.

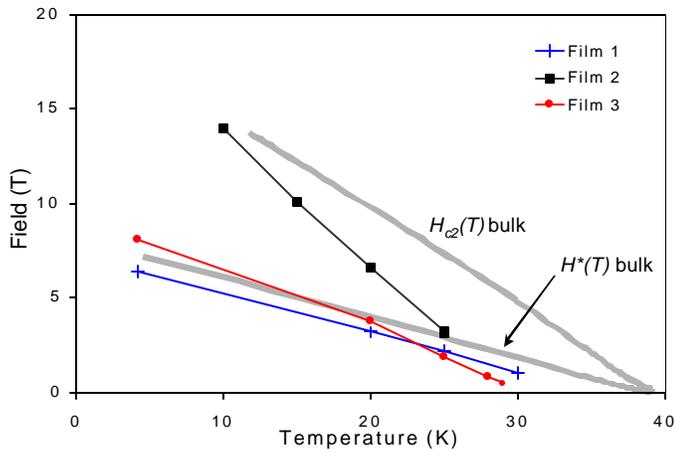

Figure 3: *H-T* phase diagram for MgB$_2$ thin films and bulk samples. Data for bulk samples are represented by the gray bands, using data from [2,4,5,8-11] for different samples and different measuring techniques. The irreversibility fields for films 1 (red) and 3 (blue) lie close to the $H^*(T)$ line for bulk materials, even though their $T_c$ values are lower than the ~39 K bulk value. The irreversibility field for film 2 increases at a faster rate of ~0.7 T/K below the critical temperature of 28.5 K, crossing the bulk $H_{c2}(T)$ line below ~10 K.

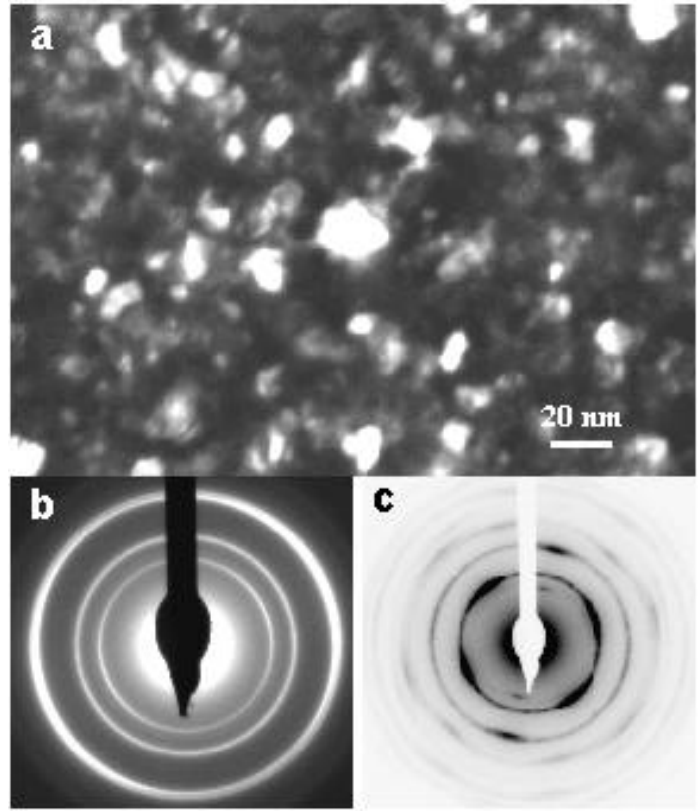

Figure 4: Transmission electron microscopy. (a) This dark-field image of a plan-view section of film 2 displays a fraction of the MgB$_2$ grains, revealing a 10 nm grain size. (b) and (c) show selected area diffraction patterns from the films. All of the rings can be attributed to either MgB$_2$ or MgO. The pattern in (b) was collected with the incident electron beam aligned approximately parallel to the film normal. It shows only complete rings, and those rings that correspond to the MgB$_2$ phase are all of the (*hk*0) type. The pattern in (c) was collected with the incident beam oriented at an angle to the film normal. It shows an uneven distribution of intensity along all MgB$_2$ rings. Considered together, the selected area diffraction patterns in (b) and (c) indicate that the film possesses a (00*l*) (i.e. *c*-axis) fiber texture oriented parallel to the film normal with little or no texture in the plane of the film.

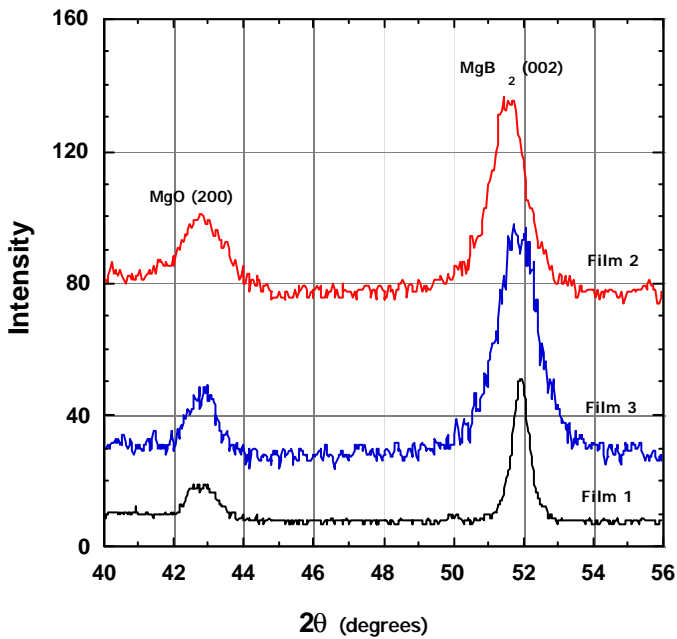

Figure 5. X-ray diffraction θ-2θ scans of films 1, 2 and 3, showing both (002) MgB$_2$ and (002) MgO peaks. The data clearly shows that the 2θ position of the MgB$_2$ (002) peak for film 2 (51.5°) is smaller than for films 1 and 3 (51.9°), which indicates the *c*-lattice parameter of film 2 is larger.

4